\pdfoutput=1
\documentclass[aps,prl,amsmath,amssymb,floatfix,lengthcheck,reprint]{revtex4-1}
\usepackage{graphicx}

\begin{document}
\title{Circuit cavity electromechanics in the strong coupling regime}
\author{J. D. Teufel$^{*}$, D. Li, M. S. Allman, K. Cicak, A. J. Sirois, J. D. Whittaker, and R. W. Simmonds}
\affiliation{National Institute of Standards and Technology, 325 Broadway, Boulder, CO 80305, USA}
\begin{abstract}
Demonstrating and exploiting the quantum nature of larger, more
macroscopic mechanical objects would help us to directly
investigate the limitations of quantum-based measurements and
quantum information protocols, as well as test long standing
questions about macroscopic quantum coherence \cite{Marshall2003}.
The field of cavity opto- and
electro-mechanics \cite{Kippenberg2008,Marquardt2009}, in which a
mechanical oscillator is parametrically coupled to an
electromagnetic resonance, provides a practical architecture for
the manipulation \cite{Braginsky1967,Dykman1978} and
detection \cite{Braginsky1975,Caves1981} of motion at the quantum
level. Reaching this quantum level requires strong coupling,
interaction timescales between the two systems that are faster
than the time it takes for energy to be
dissipated \cite{Marquardt2007,Wilson2007,Dobrindt2008}. By
incorporating a free-standing, flexible aluminum membrane into a
lumped-element superconducting resonant cavity, we have increased
the single photon coupling strength between radio-frequency
mechanical motion and resonant microwave photons by more than two
orders of magnitude beyond the current state-of-the-art. A parametric
drive tone at the difference frequency between the two resonant
systems dramatically increases the overall coupling strength. This
has allowed us to completely enter the strong coupling regime.
This is evidenced by a maximum normal mode splitting of nearly six
bare cavity line-widths. Spectroscopic measurements of these
'dressed states' are in excellent quantitative agreement with
recent theoretical predictions \cite{Agarwal2010,Weis2010}. The
basic architecture presented here provides a feasible path to ground-state cooling and subsequent coherent control and measurement of
the quantum states of mechanical motion.
\end{abstract}
\maketitle
Despite the prevalence of mechanical oscillators in many practical
technological applications, demonstrating quantum
mechanical effects in these systems has proved exceedingly
difficult. Originally cavity optomechanical
systems \cite{Kippenberg2008,Marquardt2009} were investigated as a
way to realize or even surpass the quantum limits for displacement
detection \cite{Braginsky1975,Caves1981}. More recently
experimental and theoretical progress sparked
renewed interest in these systems as a potentially viable medium for quantum
information processing. However, in order to achieve the quantum
benefits, the two disparate systems must be coupled together
strongly. Strong coupling in an analogous interaction between the
internal states of trapped ions and their motion led first to
sideband cooling to the motional ground state\cite{Diedrich1989}
and subsequently to exquisite quantum control and measurement.
Optomechanics strives to achieve this type of performance but with
a macroscopically large mechanical oscillator whose behaviour
could answer fundamental questions about the ultimate limits of
quantum based measurements and the nature of
reality \cite{Marshall2003}. Pioneering microwave electromechanical
systems utilizing parametric transducers continue their
development towards detection of gravitational
waves \cite{Braginsky1970,Linthorne1990}. Recent progress with
lithographically fabricated microwave resonators\cite{Regal2008} has enabled sideband
cooling \cite{Brown2007,Teufel2008,Rocheleau2010} and near quantum
limited detection \cite{Teufel2009,Hertzberg2010}. In
addition, the incorporation of superconducting
qubits \cite{LaHaye2009,OConnell2010} has led to the control and
measurement of a single microwave phonon \cite{OConnell2010}.

One of the most difficult challenges is engineering strong
parametric coupling between the photon cavity and the mechanical
object whose motion we are investigating. The single photon
coupling strength $g_0=G x_{\mathrm{zp}}$ is the product of
$G=d\omega_{\mathrm{c}}/dx$, the change in the photon cavity
frequency $\omega_c$ for a given displacement $x$, and the zero
point motion $x_{\mathrm{zp}}=\sqrt{\hbar/2m\Omega_{\mathrm{m}}}$
for a mechanical mode with angular frequency $\Omega_{\mathrm{m}}$
and mass $m$. This suggests designing a system using a small,
light mechanical object with a low resonant frequency (a soft
spring constant), and whose motion drastically influences the
resonant frequency of a photon cavity. Some systems have attempted
to balance all three of these preferred characteristics by using
silica microtoroids \cite{Schliesser2009}, dielectric
membranes \cite{Thompson2008}, or
nanowires \cite{Regal2008,Teufel2009,Hertzberg2010,Rocheleau2010}.
In this work, we have incorporated a free-standing aluminum
membrane into a capacitive parametric transducer to produce a
coupling strength two orders of magnitude larger than obtained
previously in a microwave circuit. This has allowed us to move entirely into a regime of
strong coupling, opening a door that leads toward ground state
cooling and subsequent quantum control and measurements with these
systems.

Imagine two parallel metal plates of area $A$ separated by a
uniform vacuum gap $d$ and electrically connected with a coiled
wire (see Fig.~1a,b). These plates form a capacitance
$C=A\epsilon_0/d$ that electrically resonates with the inductance
$L$ of the coil at an angular frequency
$\omega_{\mathrm{c}}=1/\sqrt{LC}$. When the top plate is free to
move a small distance $x\ll d$, then
$G=d\omega_{\mathrm{c}}/dx\approx-\omega_{\mathrm{c}}/(2d)$. For a
cavity with $\omega_{\mathrm{c}}/2\pi=10$ GHz and a plate
separation of $50$~nm, $G/2\pi\approx 100$~ MHz/nm. This is to be
contrasted with previous experiments that used very low-mass, high
aspect ratio
nano-wires \cite{Regal2008,Teufel2008,Teufel2009,Hertzberg2010,Rocheleau2010}.
These wires only contribute a fraction $\eta$ ($<1/1000$) of the
total capacitance so that $G=-\eta\omega_{\mathrm{c}}/(2d)$. So
while these wires have large zero-point motion, the sensitivity is
limited to $G/2\pi<100$~kHz/nm.

Our approach takes advantage of the large capacitance produced by
suspending an aluminum membrane approximately 50~nm above an
aluminum base electrode, as shown in Fig.~1b. As described later,
the sensitivity gained with this technique outweighs the reduced
$x_{\mathrm{zp}}$ due to the comparatively massive membrane
($m$=50~pg) and increases the single photon coupling
strength even beyond those achieved with optical Fabry-P\'erot
cavities \cite{Arcizet2006,Thompson2008,Groblacher2009}, where
$G\sim\omega_{\mathrm{c}}/\ell$ and $\ell$ is the length of the
optical cavity. The circuit is fabricated with wafer-scale optical
lithographic techniques developed for creating low-loss
vacuum-gap-based microwave components \cite{Cicak2010}. The nearly
circular membrane is 100~nm thick and has a diameter of 15~$\mu$m
allowing drumhead-like modes to resonate freely. The fundamental
mode is $\Omega_m/2\pi= 10.69$~MHz giving a zero point motion of
$x_{\mathrm{zp}}=4.1$~fm. The total capacitance $C\approx
38$~fF combined with a parallel inductance, $L\approx 12$~nH,
provides a single-mode microwave cavity resonance \cite{Cicak2010}
of $\omega_{\mathrm{c}}/2\pi\approx 7.5$~GHz. The device is cooled
to 40 mK, far below the superconducting transition temperature of
aluminum. To measure the motion of the membrane, we apply
microwave signals through heavily attenuated coaxial lines to the
electromagnetically coupled superconducting cavity, as shown
schematically in Fig. 1c. The transmitted signals are amplified
with a cryogenic low-noise amplifier and demodulated at room
temperature with either a commercial vector network analyzer (for
characterizing the cavity mode) or a spectrum analyzer (for
characterizing the mechanical mode).

Fig.~2a shows the magnitude of transmission, $|T|^2$, near the
cavity resonance at sufficiently low microwave power that
radiation pressure effects can be neglected. A Lorentzian fit
yields a resonance frequency of
$\omega_{\mathrm{c}}/2\pi=$7.47~GHz and a loaded intensity decay
rate of $\kappa/2\pi=$170~kHz.  The depth of the dip at resonance
shows that the circuit is overcoupled, so that the dominant source
of damping is the intentional inductive coupling to the
transmission line, $\kappa_{\mathrm{ex}}/2\pi=$130~kHz, which is
much greater than the intrinsic decay rate,
$\kappa_{\mathrm{0}}/2\pi=$40~kHz. The motion of the drumhead mode
modulates the capacitance and thus the frequency of the electrical
resonator, creating sidebands above and below the microwave drive
frequency at $\omega_{\mathrm{d}}\pm\Omega_{\mathrm{m}}$. Fig.~2b
shows the noise power of the upper sideband due to the thermal
motion of the drum at its fundamental mode,
$\Omega_{\mathrm{m}}/2\pi=10.69$~MHz. These data show that the
mechanical resonance has an intrinsic damping rate of
$\Gamma_{\mathrm{m}}/2\pi=30$~Hz and a high mechanical quality
factor $Q_{\mathrm{m}}=360,000$, consistent with the tensile
stress generated by thermal contractions upon
cooling \cite{Regal2008,Teufel2008,Teufel2009}.  This system is in
the resolved sideband regime, in which the mechanical resonance
frequency is much larger than the cavity linewidth,
$\Omega_{\mathrm{m}}/\kappa=63$.  This is a prerequisite for
sideband cooling to the ground
state \cite{Marquardt2007,Wilson2007} and for observing normal-mode
splitting in the driven optomechanical
system \cite{Marquardt2007,Dobrindt2008}.

The quantum-mechanical behavior of this parametrically coupled
system is described by the interaction Hamiltonian:
$H_{\mathrm{I}}=-\hbar a^\dagger a g_0 (b^\dagger+ b)$, where
$a^{\dagger}$ and $b^{\dagger}$ are the creation operators for
photons and phonons, respectively. Because the motion of this drum
strongly influences $\omega_{\mathrm{c}}$, microwave signals can
be used not only to detect the motion of the oscillator but also
to impart backaction forces on it. The radiation pressure force of
the microwave drive photons gives rise to ``optical" spring and
damping effects \cite{Braginsky1967,Dykman1978}. The interaction
Hamiltonian can be linearized in a frame co-rotating with the
drive, taking the form: $H_{\mathrm{I}}=- \hbar g
(a^{\dagger}+a)(b^{\dagger}+b)$, where
$g=g_0\sqrt{n_{\mathrm{d}}}$ is the linearized optomechanical
coupling rate \cite{Marquardt2007,Wilson2007,Dobrindt2008}, and
$n_{\mathrm{d}}$ is the number of drive photons in the cavity.  If
the drive is detuned so that its upper mechanical sideband is near the cavity resonance, $\delta=(\omega_{\mathrm{d}}+\Omega_{\mathrm{m}})-\omega_{\mathrm{c}}\ll\Omega_{\mathrm{m}}$
as shown in Fig 2c, the modified mechanical resonance frequency
$\Omega'_{\mathrm{m}}$ and damping rate $\Gamma'_{\mathrm{m}}$
closely follow the imaginary and real parts of the cavity
response. In the resolved sideband regime these quantities are
well approximated by \cite{Marquardt2007,Wilson2007}:
\begin{equation}
\Omega'_{\mathrm{m}} \approx \Omega_{\mathrm{m}}+\frac{4g^2
\delta
}{\kappa^2+4\delta^2},
\end{equation}
\begin{equation}
\Gamma'_{\mathrm{m}} \approx \Gamma_{\mathrm{m}}+\frac{4g^2
\kappa}{\kappa^2+4\delta^2}.
\end{equation}

Fig.~2d shows the measured effects of this dynamical backaction on
the drum as a function of $\delta$.  The incident microwave
power $P_{\mathrm{in}}$ is held constant at 10~pW. As this power
is applied very far from the cavity resonance, it results in a
greatly reduced number of photons in the cavity, given by
$n_{\mathrm{d}}=2P_{\mathrm{in}}\kappa_{\mathrm{ex}}/\hbar
\omega_{\mathrm{d}}(\kappa^2+4\Delta^2)$, where $\Delta=\omega_{\mathrm{d}}-\omega_{\mathrm{c}}$. Even for this moderate
power microwave drive with $n_{\mathrm{d}}\leq 800$, the effects
on the mechanical oscillator are quite striking; the intrinsic
mechanical damping is dominated by the damping from the microwave
photons. Fitting these results to Eqs.~1 and 2 (shown in black)
gives $G/2\pi$=56~MHz/nm while showing
overall excellent agreement with the theoretical predictions.

Just as the microwave photons strongly affect the mechanical
mode, the symmetry of the parametric interaction suggests that
the mechanics should influence the cavity mode.  To
investigate this, we apply both a microwave drive tone
$\omega_\mathrm{d}$ and a second probe tone $\omega_{\mathrm{p}}$.
Here the drive tone will induce an interaction between the
mechanics and the cavity, while the probe tone will monitor the
response of the cavity. This technique provides a way to measure
the spectroscopy of the ``dressed" cavity states in the presence
of the electromechanical interaction. Fig. 3a shows a series of
cavity probe spectra taken with successively higher microwave power
applied with $\Delta = -\Omega_{m}$.  Once the drive power is
large enough that $g\sim
\sqrt{\Gamma_{\mathrm{m}}\kappa}$, the mechanical sideband of the
driving field appears in the cavity response function.  As
$n_{\mathrm{d}}$, and hence $g$,  increase so
does the normalized probe transmission at the cavity resonance,
$|T(\omega_{\mathrm{c}})|$. The width of this peak also increases
and is given by the modified mechanical damping rate in Eq.~2. This electromechanical effect can be understood as the result of a radiation
pressure force at the beat frequency between the drive and probe
photons, which drives the motion of the drum near its resonance
frequency. The driven motion results in a mechanical sideband on
the the drive field that can interfere with the probe field and
hence a modified probe spectrum. This interference is the
mechanical analogue of electromagnetically induced
transparency \cite{Boller1991} well known in atomic physics, and
has only recently been addressed in the context of optomechanics,
both theoretically \cite{Agarwal2010,Weis2010} and
experimentally \cite{Weis2010}.  Applying this theory to our
electrical circuit  implies that the transmission spectrum is
\begin{equation}
T=1-\frac{\kappa_{\mathrm{ex}}\left(1-j \chi\right)}{\kappa+2 j
\left(\omega_{\mathrm{p}}-\omega_{\mathrm{c}}\right)+4 \chi
\left(\omega_{\mathrm{d}}-\omega_{\mathrm{c}}\right)},
\end{equation}
where
\footnotesize
\begin{equation*}
\chi=\frac{4g^2\Omega_{\mathrm{m}}}{\left[\kappa+2 j
\left(\omega_{\mathrm{p}}-2\omega_{\mathrm{d}}+\omega_{\mathrm{c}}\right)\right][\Omega_{\mathrm{m}}^2-\left(\omega_{\mathrm{p}}-\omega_{\mathrm{d}}\right)^2+j\left(\omega_{\mathrm{p}}-\omega_{\mathrm{d}}\right)\Gamma_{\mathrm{m}}]}\nonumber.
\end{equation*}
\normalsize
At high enough power, $\Gamma'_{\mathrm{m}}$  becomes comparable
to or greater than $\kappa$, at which point Eqs. 1 and 2, are no
longer valid.  This is precisely the point at which the driven system
enters the strong coupling regime, where the coupling exceeds the
intrinsic dissipation in either of the original modes ($g\geq
\kappa \gg \Gamma_{\mathrm{m}}$).  The eigenmodes of the driven,
coupled system are now hybrids of the original radio-frequency
mechanical and microwave electrical resonances.  The system
exhibits the well-known normal-mode splitting of two strongly
coupled harmonic oscillators, as was recently demonstrated in a
room-temperature optomechanical system \cite{Groblacher2009}. For our
device, progression into the strong coupling regime is shown in
Fig.~3a,b with a crossover occurring at $n_{\mathrm{d}}\sim 10^5$.
In this regime, the damping rate of each of the two normal modes approaches
$(\kappa+\Gamma_{\mathrm{m}})/2$, and the coupling results in a
splitting of $2g$.  In Fig.~3c, $g$ is extracted by fitting each
spectrum at a given drive power to Eq.~3. The splitting shows the
expected $\sqrt{n_{\mathrm{d}}}$ dependence with a single photon
coupling rate of $g_0/\pi= 460$~Hz. At the highest drive power,
$n_{\mathrm{d}}=5\times 10^6$, the splitting is $g/\pi=1$~MHz,
exceeding both $\kappa$ and $\Gamma_{\mathrm{m}}$.

By measuring the in-phase and quadrature components of the
transmitted probe signal, we determine the real and imaginary
parts of the cavity spectrum in the presence of the
electromechanical interactions. We find excellent agreement
between theory and experiment for both magnitude and phase over
the entire range of accessible detunings and powers. Fig~4a shows
the magnitude and phase of the probe transmission for three
arbitrary values of $\delta$ when $g/\pi=150$~kHz. The black lines
are fits to the magnitude and argument of Eq.~3. Two-dimensional
plots of $|T|$ are shown in Fig.~4 as a function of both
$\omega_{d}$ and $\omega_{p}$.  At low drive amplitude (Fig.~4b),
the narrow mechanical sideband appears as a sharp dip or peak in
transmission whenever
$\omega_{p}=\omega_{d}+\Omega_{\mathrm{m}}'$.  As the drive
amplitude is increased (Fig. 4c), the dispersive coupling between
the normal modes becomes apparent. While the mechanical sideband
is narrow when it is far from resonance, as it approaches
$\omega_{\mathrm{c}}$, it inherits the cavity's larger damping
rate.  At the largest coupling, the normal mode splitting is
appreciable for a broad range of drive and probe frequencies as
shown in Fig. 4d.

Assuming thermal equilibrium with the cryostat temperature of 40
mK, the thermal population of the microwave cavity
$n_{\mathrm{c}}$ and the mechanical mode $n_{\mathrm{m}}$ would be
$10^{-4}$ and 80 quanta, respectively. Although the measurements
shown here do not distinguish between classical or quantum
behavior, this system possesses the coupling strength
and low thermal decoherence rates necessary to realize quantum entanglement
between the mechanical and electromagnetic modes. With the cavity
initially in its quantum-mechanical ground state, the strong
coupling implies that resolved sideband cooling can be used to
reduce the occupancy of the mechanical mode by a factor of
$\kappa/\Gamma_{\mathrm{m}}\approx5000$ \cite{Marquardt2007,Wilson2007,Dobrindt2008}, placing both
(coupled) resonators in their ground state.  This is the quantum-enabled regime in which the thermal decoherence rate is
$\Gamma_{\mathrm{th}}=\bar{n}_{\mathrm{m}}\Gamma_{\mathrm{m}}\ll
\kappa \ll\Omega_{\mathrm{m}}$.  Strongly coupled quantum harmonic
oscillators pave the way for many future experiments, including
quantum information processing and storage, and the generation of
nonclassical states of both the photon and phonon fields
\cite{Kippenberg2008,Marquardt2009,Akram2010}.

This experiment demonstrates the greatly improved
electromechanical coupling that is possible by engineering a
microwave resonant circuit to include a micromechanical membrane.
We have shown quantitative agreement with the theoretical
predictions for dynamical backaction. The parametric interaction
allows for strong coupling between two harmonic modes, at $\sim11$
MHz and $\sim7.5$ GHz, far off resonance with each other, with an
energy transfer rate much faster than the energies decays from
either system. Just as electromagnetically induced transparency
enabled such innovations as slow light and photon storage, the
analogous electromechanical effects demonstrated here should already give
rise to group delay in microwave signals of
$\Gamma_{\mathrm{m}}^{-1} \approx$5~ms and storage of microwave
quantum states on the timescale of $\Gamma_{\mathrm{th}}^{-1}\approx$
70~$\mu$s \cite{Weis2010}. Future experiments, incorporating an
even lower noise microwave amplifier
\cite{Castellanos-Beltran2008,Teufel2009} into the measurement,
will facilitate measurements of the thermal population at the
single-phonon and single-photon level, and could allow
Heisenberg-limited detection of displacement or force with
sensitivities at the attometer and attonewton level. Lastly, straightforward integration of this electromechanical circuit with
superconducting qubits \cite{Cicak2010} will lead to
generation and manipulation of motional quantum states \cite{OConnell2010}.
\section{Acknowledgements} We thank A. W. Sanders for taking the
micrograph in Fig.~1b, and gratefully acknowledge discussions with
T. Donner, J. H. Harlow and K. W. Lehnert. This work was
financially supported by NIST and DTO. Contribution of the U.S.
government, not subject to copyright. Correspondence should be addressed to J.D.T (john.teufel@nist.gov).
%

\pagebreak

\begin{figure*}[h]
\includegraphics{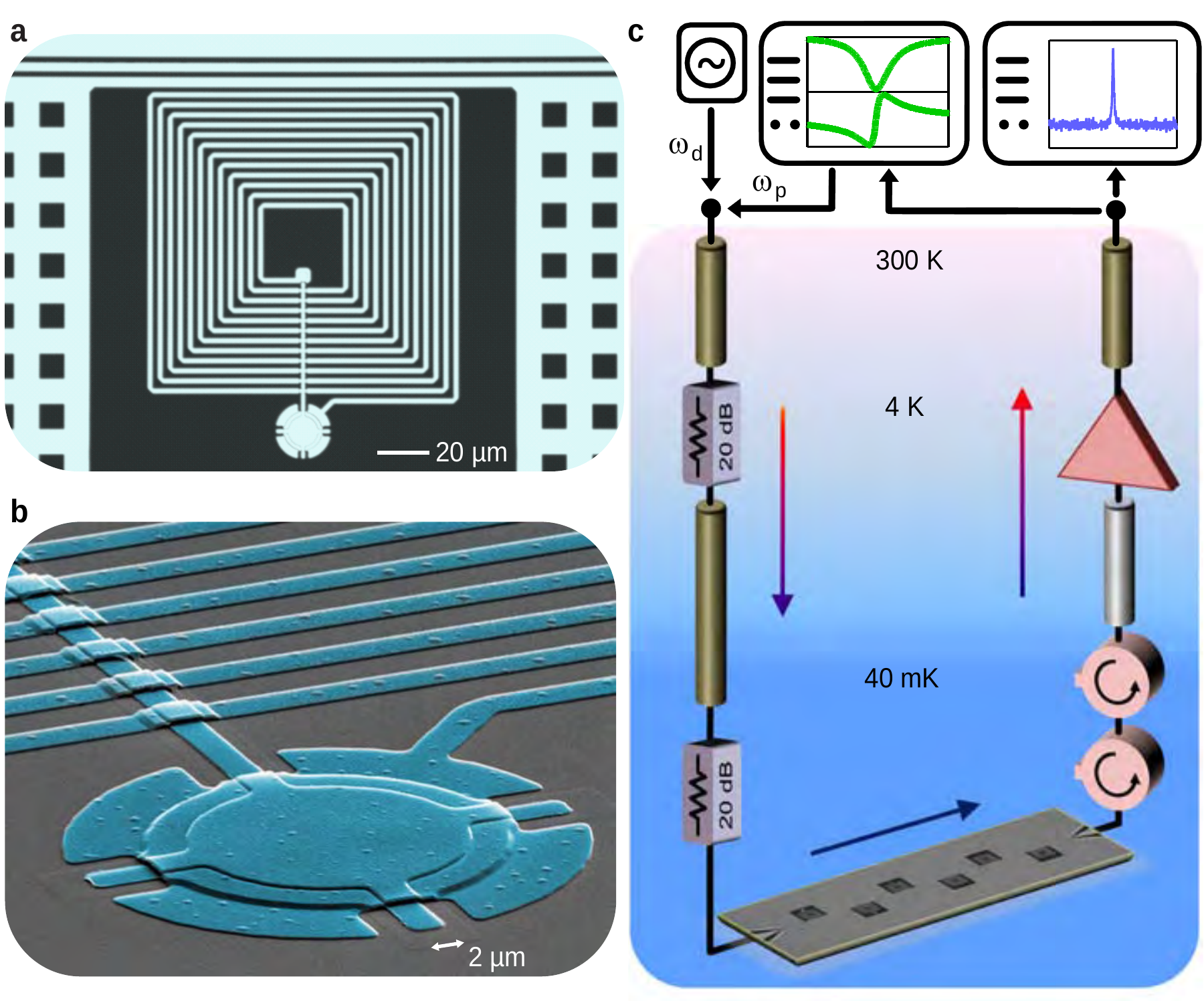} 
\caption{\textbf{Schematic description of the experiment.}
\textbf{a}, Colorized optical micrograph of the microwave
resonator formed by a spiral inductor shunted by a parallel-plate
capacitor.   \textbf{b}, Colorized scanning electron micrograph
shows the upper plate of the capacitor is suspended  $\sim$50~nm
above the lower plate and is free to vibrate like a taught,
circular drum.  The metallization is sputtered aluminum (blue)
patterned on a sapphire substrate (black). \textbf{c}, This
circuit is measured by applying microwave signals near the
electrical resonance frequency through resistive coaxial lines.
The outgoing signals, in which the mechanical motion is encoded as
modulation sidebands of the applied microwave tone, is coupled to
a low noise, cryogenic amplifier via a superconducting coaxial
cable. Cryogenic attenuators on the input line and isolators on
the output line ensure that thermal noise is reduced below the
vacuum noise at microwave frequencies.}
\end{figure*}
\pagebreak
\begin{figure*}[h] 
\includegraphics{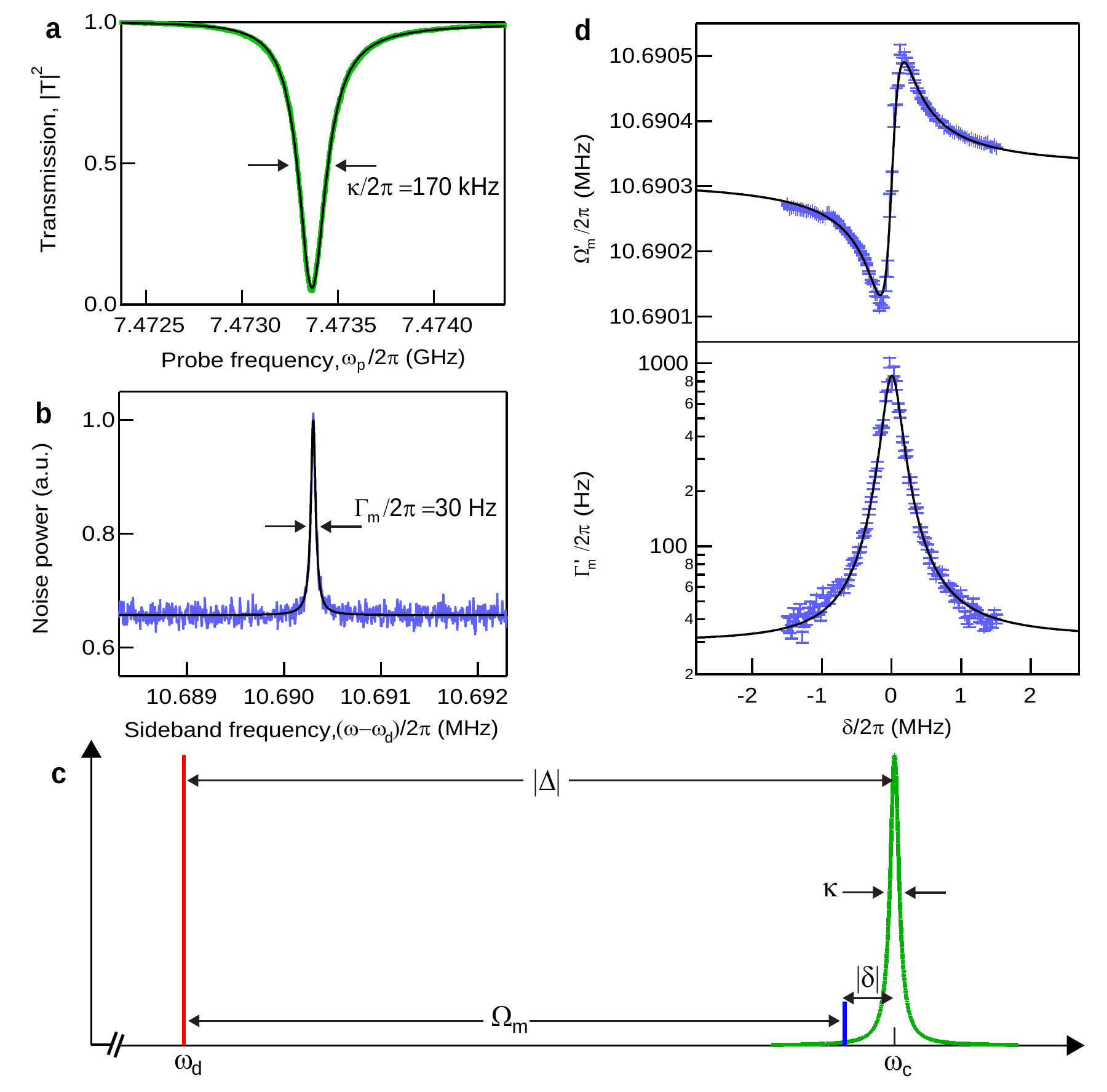}
\caption{  \textbf{Characterization of mechanical and microwave
resonances.} \textbf{a}, Measured probe transmission spectrum
(green) and Lorentzian fit (black) of the microwave circuit at low
power, where optomechanical effects are negligible.  The width of
the resonance yields the overcoupled,  intensity decay rate
$\kappa/2\pi$=170~kHz. \textbf{b}, The mechanical resonance
manifests itself as a peak in the noise spectrum (blue), which
appears $\Omega_{\mathrm{m}}$ above and below the microwave drive frequency,
due to the thermal motion of the drum up- or down-converting
microwave photons.  At low microwave power, where backaction
effects are negligible, the Lorentzian fit (black) yields an
intrinsic mechanical dissipation rate
$\Gamma_{\mathrm{m}}/2\pi$=30~Hz ($Q_{\mathrm{m}}$=360,000).
\textbf{c},  Schematic diagram for the relative frequencies of the
microwave drive (red) and the upper mechanical sideband (blue)
with respect to the narrow cavity resonance (green). \textbf{d},
The modified mechanical resonance frequency $\Omega'_{\mathrm{m}}$
and damping rate $\Gamma'_{\mathrm{m}}$ as a function of the
relative detuning  $\delta$  fit well to the theory of dynamical
backaction (black), yielding $G/2\pi$=56~MHz/nm.}
\end{figure*}
\pagebreak
\begin{figure*}[h] 
\includegraphics{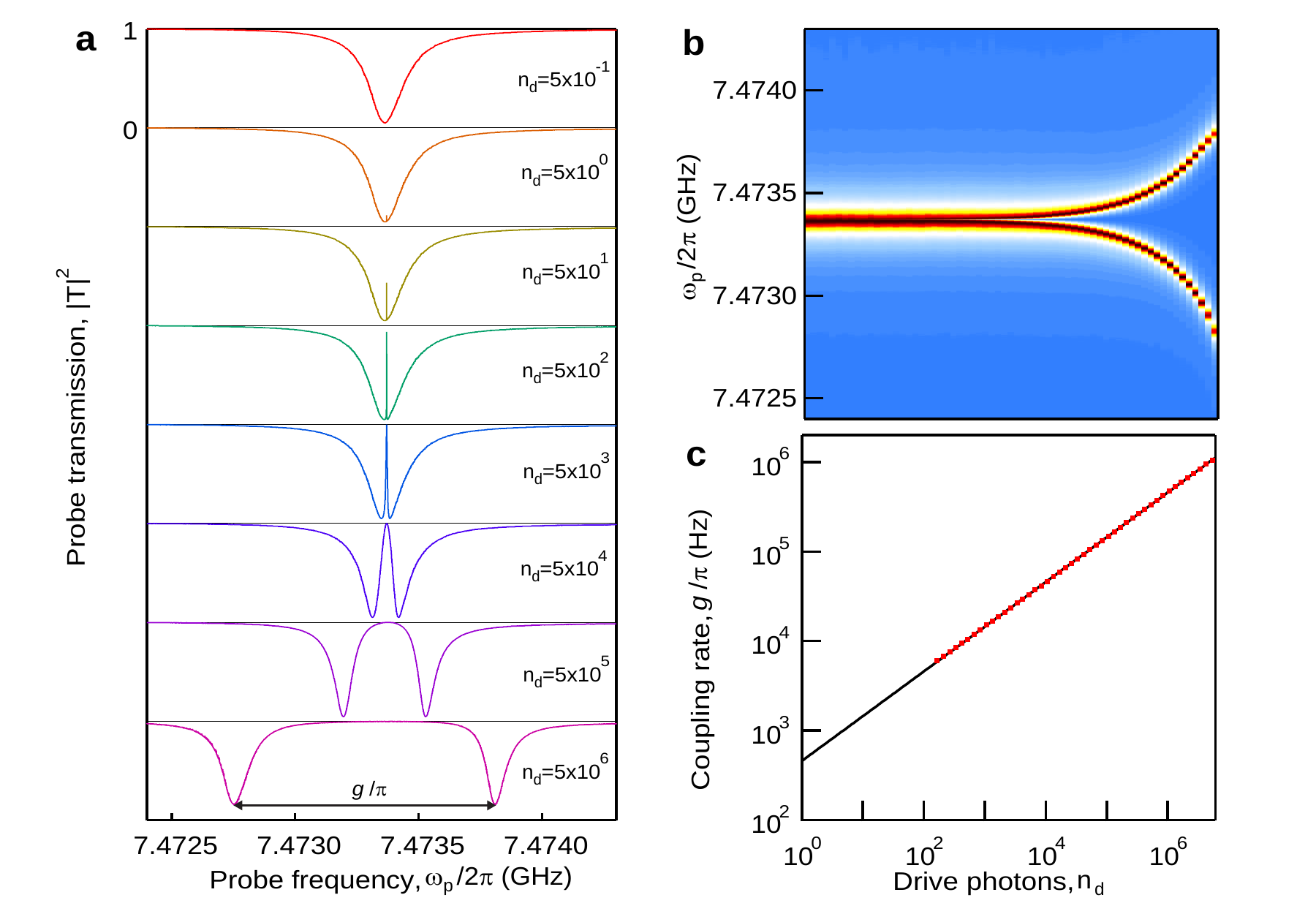}
\caption{ \textbf{Demonstration of the strong-coupling regime.}
\textbf{a}, Normalized microwave cavity transmission  in the
presence of a microwave drive applied with $\Delta = -\Omega_{m}$,
with successive plots for increasing drive amplitude
$n_{\mathrm{d}}$.  At moderate drive amplitude ($n_{\mathrm{d}}
\approx 10$), the interference between the drive and probe photons
results in a narrow peak in the cavity spectrum, whose width is
given by the mechanical linewidth $\Gamma'_{\mathrm{m}}$.  When
$\Gamma'_{\mathrm{m}}$ becomes comparable to $\kappa$, the cavity
resonance splits into normal modes.  The eigenmodes of the system
are no longer purely electrical or mechanical, but are a pair of
hybrid electromechanical resonances.  \textbf{b},  The
transmission as a function of probe frequency and number of drive
photons shows that the driven system enters the strong coupling
regime ($g\geq \kappa,\Gamma_{\mathrm{m}}$). Here, the logarithimic color scale show the transmission as it varies from -13~dB (dark red) to 0~dB (blue). \textbf{c}, The
measured coupling rate $g$ (red) follows the expected dependence
on the number of drive photons, with a fit to
$\sqrt{n_{\mathrm{d}}}$ shown in black. }
\end{figure*}
\pagebreak
\begin{figure*}[h] 
\includegraphics{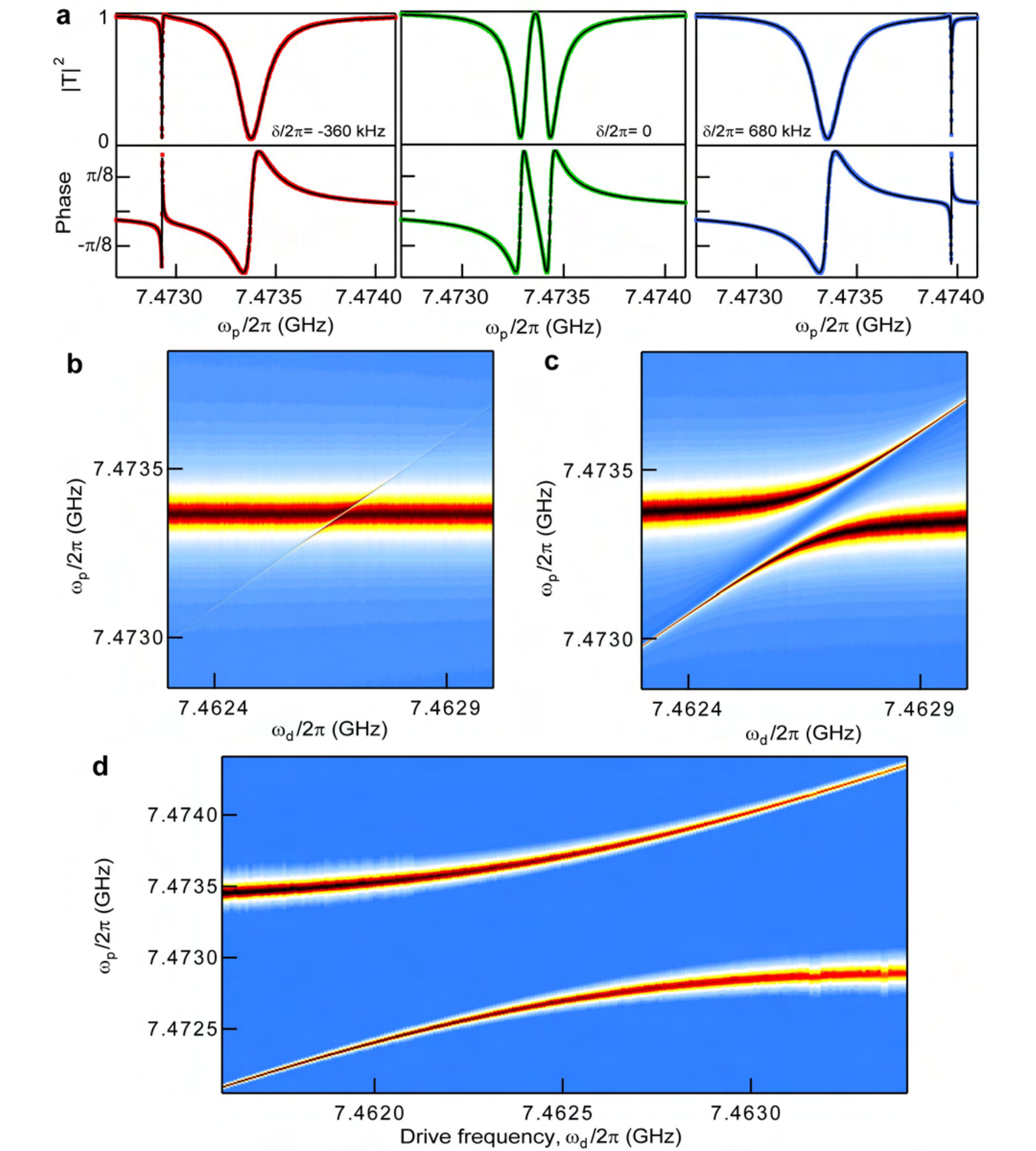}
\caption{\textbf{Spectroscopy in the strong-coupled regime.}
\textbf{a}, The normalized magnitude and phase of the cavity
transmission in the presence of a strong microwave drive.  The
data shown in red, green and blue are for three different detunings $\delta$
with fits to Eq. 3 (black). \textbf{b},  For
relatively low drive power ($n_{\mathrm{d}} \approx 10$),  the
mechanical sideband appears as a sharp dip in the transmission at
a frequency $\omega_{\mathrm{d}}+\Omega'_{\mathrm{m}}$.
\textbf{c},  When the drive amplitude is increased
($n_{\mathrm{d}} \approx 10^4$),  the two resonances show an
avoided crossing between the eigenmodes of the coupled system.
\textbf{d},   For the largest amplitude drive ($n_{\mathrm{d}}
\approx 5 \times 10^6$), the driven transmission spectra are split
by much more than $\kappa$ or $\Gamma_{\mathrm{m}}$ for a broad
range of detunings.}
\end{figure*}

\end{document}